\documentclass[twocolumn]{revtex4-1}
\usepackage{graphicx}
\usepackage{epstopdf}
\usepackage{amsmath}
\begin{document}
\title{Net-proton number kurtosis and skewness in nuclear collisions:\\
Influence of deuteron formation}
\author{Zuzana Feckov\' a$^{a,b}$}
\author{Jan Steinheimer$^{c}$}
\author{Boris Tom\'a\v{s}ik$^{b,e}$}
\author{Marcus Bleicher$^{c,d}$}
\affiliation{$^a$Univerzita Pavla Jozefa \v{S}af\'arika, 
\v{S}rob\'{a}rova 2, 04001 Ko\v{s}ice, Slovakia}
\affiliation{$^b$Univerzita Mateja Bela, Tajovsk\'eho 40, 97401 Bansk\'a Bystrica, Slovakia}
\affiliation{$^c$Frankfurt Institude for Advanced Studies, 
Johann Wolfgang Goethe-Universit\"at, Ruth-Moufang-Strasse 1, 60438 Frankfurt am Main}
\affiliation{$^d$Institut f\"ur Theoretische Physik, 
Johann Wolfgang Goethe-Universit\"at, Max-von-Laue-Strasse 1, 60438 Frankfurt am Main}
\affiliation{$^e$\v{C}esk\'e vysok\'e u\v{c}en\'i technick\'e v Praze, 
FJFI, B\v{r}ehov\'a 7, 11519 Praha 1, Czech Republic}

\keywords{kurtosis, deuterons, quark-gluon-plasma, heavy ion collisions, RHIC-BES}
\begin{abstract}
We explore the influence of deuteron formation in the late stage of nucleus-nucleus reactions on the fluctuations observed in the final net-proton yields around midrapidity. At each investigated energy, the produced
(anti-)proton yield at chemical freeze-out is assumed to fluctuate according to a Poisson distribution and in each event the probability for deuteron formation by coalescence is 
proportional to $(dN_{\mathrm{proton}}/dy)^2$. 
The protons that are then clustered in deuterons are 
usually not included in the experimental measurement of the net-proton fluctuations,
therefore, we subtract these clustered protons from the final state proton number for the calculation of the net-proton fluctuations (the same is done in the anti-proton sector). Due to the non-linear deuteron formation probability the resulting distribution is not a Skellam distribution, but shows the interesting feature of a decrease in the kurtosis $\kappa \sigma^2$ and a local maximum in the skewness $S \sigma$ observables as the collision energy decreases. 
\end{abstract}
\maketitle


\section{Introduction}
The exploration of the properties of ultra-hot and super-dense nuclear matter is among the most challenging scientific endeavors today. The collisions of heavy ions allow to explore these properties in the laboratory for the first time with unprecedented precision. Fluctuation observables have gained very great interest in recent years
as they may be sensitive to the structure of the QCD phase diagram. In particular they can serve as an indicator for the first order deconfinement phase transition and the associated critical end point \cite{Sasaki:2007db,Randrup:2009gp,Steinheimer:2012gc,Stephanov:1999zu}. Particle number fluctuations were first explored in experiments at the SPS \cite{Alt:2008ab,Anticic:2008aa}. With the availability
of the first lattice QCD data on higher order susceptibilities in \cite{Allton:2005gk} these investigations \cite{Stephanov:1998dy,Stephanov:1999zu,Jeon:2000wg,Asakawa:2001mn,Koch:2001zn,Koch:2008zz,Karsch:2010ck} have moved to the next level and have been intensively explored with STAR's 
Beam Energy Scan (BES)
at the RHIC facility \cite{Aggarwal:2010wy}. Recent results from the BES program indicate a 
non-trivial beam dependence of the higher order net-baryon number cumulants.
This is of particular interest, as these higher order cumulants are very sensitive to the existence of a second order phase transition \cite{Stephanov:2008qz,Koch:2008ia} and the associated fluctuations and correlations.

While irregular structures have been observed in the experimental data, the interpretation and final experimental validation are still missing today. For example,
 effects of conservation laws \cite{Schuster:2009jv,Bzdak:2012an,Sakaida:2014pya}, final state hadronic interactions \cite{Kitazawa:2011wh}, repulsive hadronic interactions \cite{Begun:2012rf} and acceptance effects \cite{Bzdak:2012ab} as well as efficiency corrections \cite{Bzdak:2013pha} are considered to be very important for the interpretation of experimental results.

It is therefore essential to understand all trivial and non-trivial effects which have an influence on the observed cumulant ratios, in order to make conclusions on the existence and discovery of a phase transition and critical endpoint.

In the following we suggest a further uncertainty, arising from nuclear cluster formation, that needs to be taken into account before final conclusions from these fluctuation observables should be drawn.


\section{Set-up}
For this exploratory study, we will stay with a simple physical picture to elucidate the main effect of cluster production on the observable fluctuations. It is clear that more elaborate studies will be needed to allow for a detailed and more quantitative investigation of the effect of cluster formation.

The scenario is as follows. We assume that the number of chemically frozen out protons 
in some midrapidity interval $n_i$ fluctuates according to a
Poisson distribution
\begin{equation}
P_i(n_i)=\lambda_p^{n_i} \frac{e^{-\lambda_p}}{n_i!} \, .
\end{equation}
Here, $\lambda_p$ is the mean number of these \emph{initial} protons. It is important 
to realize that after deuteron coalescence some of them will be hidden in deuterons and 
thus will disappear when counting the observed final state protons.   Therefore,
$\lambda_p$ is unknown at this point but will be determined from the \emph{observed}
proton and deuteron yields later. 
At higher collision energies a similar procedure should be done for antiprotons. 
We will show, however, 
that for the energy range of SPS and below their influence can be neglected.
 
\begin{figure}[t]
\includegraphics[width=0.5\textwidth]{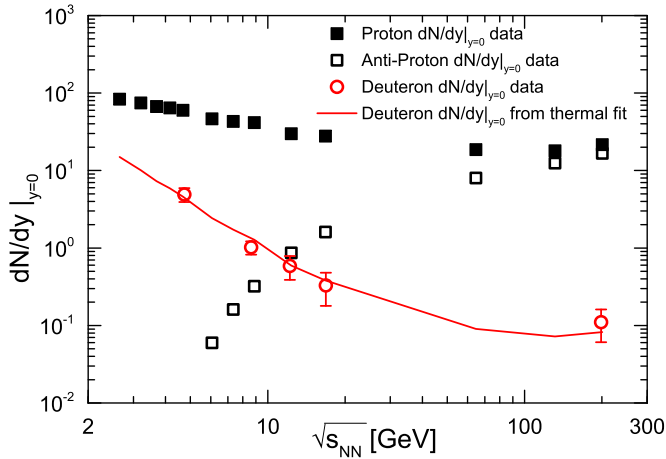}
\caption{\label{fig:deut} The midrapidity proton, antiproton and deuteron yields from experimental data (symbols) and the corresponding deuteron yield (red solid line) which was extracted using the thermal fit to the $d/p$ ratio, for central Au+Au reactions.}
\end{figure} 

Next, we assume that the probability to form a deuteron by coalescence after the kinetic 
freeze-out is in each event proportional to the actual initial proton yield squared in 
that event \cite{Butler:1963pp,Nagle:1996vp}, i.e.
\begin{equation}\label{coal}
\lambda_d = B n_i^2\, , 
\end{equation}
with a coalescence parameter $B$ which depends only on the beam energy. At this point we make the assumption that the
neutron yield in a single event is proportional to the proton yield in that event. If 
the neutron yield was uncorrelated with the proton number the correlation between proton and deuteron production
in a single event is 
weakened. This feature may be experimentally checked by measuring the production
yield of deuterons as a function of the proton number, in a specific centrality bin.
Further, it is important to note, that this assumption is only valid in the absence of
volume fluctuations. In a scenario where volume fluctuations dominate the observed proton number fluctuations
(due to e.g. wide centrality bin selections), both the proton and deuteron number fluctuations will
scale with the fluctuations of the volume $n_i \propto n_d \propto V$. As a result the measured proton number
distribution will be described by a Skellam distribution because proton and deuteron fluctuations can be described with independent Poisson distributions.
In the case of vanishing volume fluctuations, as is usually ensured by the tight cuts perdormed by the experiments, the deuteron number can be considered
correlated with the initial proton number, as presented in eq.~(\ref{coal}).
 
The deuteron yield $n_d$ in an event with initial proton multiplicity $n_i$ then fluctuates according
to a Poisson distribution
\begin{equation}
P_d(n_d|n_i)=\lambda_d^{n_d} \frac{e^{-\lambda_d}}{n_d!} =
\left ( B n_i^2 \right )^{n_d} \frac{e^{-B n_i^2}}{n_d!} 
\, .
\end{equation} 
The observed deuteron distribution is obtained by summing over initial proton 
yields as
\begin{equation}
P_d(n_d) = \sum_{n_i} P_d(n_d|n_i) P_i(n_i)\, .
\label{e:ddist}
\end{equation}

To get the observed proton number, we have to subtract from the initial protons those which 
went into deuterons
\begin{equation}
\frac{dN_{\mathrm{proton}}}{dy} = n_p = n_i - n_d\, .
\label{e:pdiff}
\end{equation}
The distribution of $n_p$ then follows from the appropriate convolution of $P_i(n_i)$ with
$P_d(n_d|n_i)$
\begin{multline}
P(n_p) = \sum_{n_i\ge n_p}\sum_{n_d} P_i(n_i) P_d(n_d|n_i) \delta_{n_p,n_i-n_d}\\
= \sum_{n_i\ge n_p} P_i(n_i) P_d(n_i-n_p|n_i)\, .
\label{e:pdist}
\end{multline}

The two parameters in this model are the mean \emph{initial} proton number $\lambda_p$
and the coalescence factor $B$. They can be fixed from the mean \emph{observed}
proton multiplicity and the mean observed deuteron multiplicity
\begin{eqnarray}
\langle n_p \rangle & = & \sum_{n_p} n_p P(n_p) \\
\langle n_d \rangle & = & \sum_{n_d} n_d P_d(n_d)\, ,
\end{eqnarray}
where $P(n_p)$ and $P_d(n_d)$ are determined in eqs.~(\ref{e:pdist}) and (\ref{e:ddist}). 

Then, the distribution $P(n_p)$ 
can be used to calculate any moments of the proton distribution. 

As the data on deuteron multiplicities for various collision energies are scarce, we note that 
those that are available are actually well fitted by the thermal model. The deuteron-to-proton
multiplicity ratio at midrapidity can be parametrized as
\[
\frac{d}{p} = 0.8 \left [ \frac{\sqrt{s_{NN}}}{1\, \mathrm{GeV}} \right ]^{-1.55} + 0.0036\, , 
\]
as shown in \cite{Andronic:2005yp}. Figure \ref{fig:deut} depicts the midrapidity proton anti-proton and deuteron yields from experimental data (see \cite{Andronic:2005yp} and references therein) and the thermal model parametrization of the deuteron yield \cite{Andronic:2010qu} which serve as input for our calculations.

Using this input we calculate the variance $\sigma$, 
skewness $S$ and the kurtosis $\kappa$ from the distribution $P(n_p)$
\begin{eqnarray}
\sigma^2 & = & \left \langle (n_p - \langle n_p\rangle )^2 \right \rangle\\
S & = & \frac{\left \langle( n_p - \langle n_p\rangle )^3 \right \rangle}{\sigma^3}\\
\kappa & = & \frac{\left \langle ( n_p - \langle n_p\rangle )^4 \right \rangle}{\sigma^4} -3\, .
\end{eqnarray}

\section{Analytic Results}

\begin{figure}[t]
\includegraphics[width=0.5\textwidth]{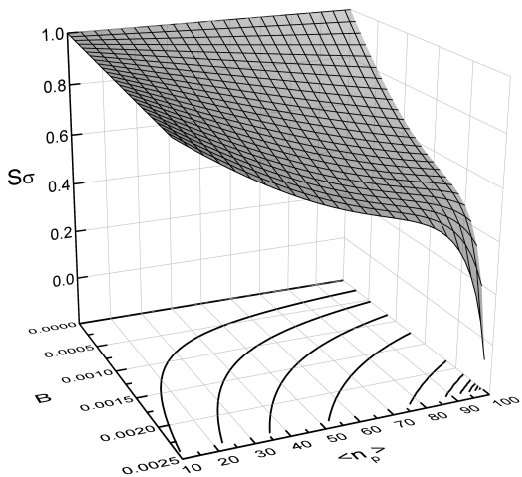}
\caption{\label{f:sk} 
Scaled skewness $S\sigma$ as a function of the mean observed proton number and 
the coalescence parameter $B$ in the region of values relevant for AGS and SPS collision
energies.}
\end{figure}
\begin{figure}[t]
\includegraphics[width=0.5\textwidth]{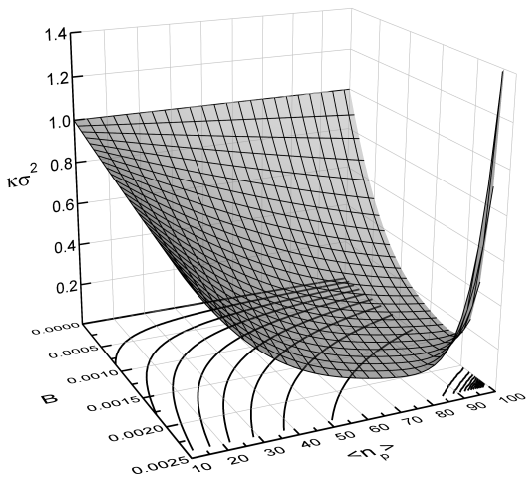}
\caption{\label{f:ku} 
Same as Fig.~\ref{f:sk} but for the scaled kurtosis  $\kappa\sigma^2$.}
\end{figure}

For illustration, in Fig.~\ref{f:sk} we show the scaled skewness $S\sigma$ 
and in Fig.~\ref{f:ku} the scaled kurtosis $\kappa\sigma^2$ 
of the distribution $P(n_p)$ as functions of the observed mean proton number 
$\langle n_p \rangle$ and the coalescence parameter $B$. 

Recall that both $S\sigma$ and $\kappa\sigma^2$ would be equal to unity if the underlying observed proton distribution 
was Poissonian.
(There is  deviation from 1 for $S \sigma$ if the distribution was 
Skellam, but we will scale it out later in the presentation of results.)
 This is really so in the absence of deuterons, i.e.~when $B=0$ or for small numbers of
observed protons.

We also see a strong deviation from unity for large values of $\langle n_p \rangle$ and $B$.
In fact the scaled kurtosis may take values which are both 
smaller or larger than unity.

For a medium value of proton multiplicity and/or coalescence parameter 
the obtained deuteron number limits the possible observed proton number 
fluctuation. Since the mean deuteron number scales with the square of 
proton number, if the proton number becomes large in an event, the number of deuterons 
makes even larger relative deviation from the mean.
This limits the fluctuations of the observed protons mainly on the upper side of the 
distribution and the tails of the distribution are below a comparable Poisson 
distribution. When the deuteron number grows further, it starts pushing 
the observed proton number down (in agreement with eq.~(\ref{e:pdiff})). This 
increases the deviations of the proton number from the mean to lower values. The left tail of 
the distribution is above Poissonian and the kurtosis grows. The skewness becomes 
clearly smaller. 

\begin{figure}[t]
\includegraphics[width=0.5\textwidth]{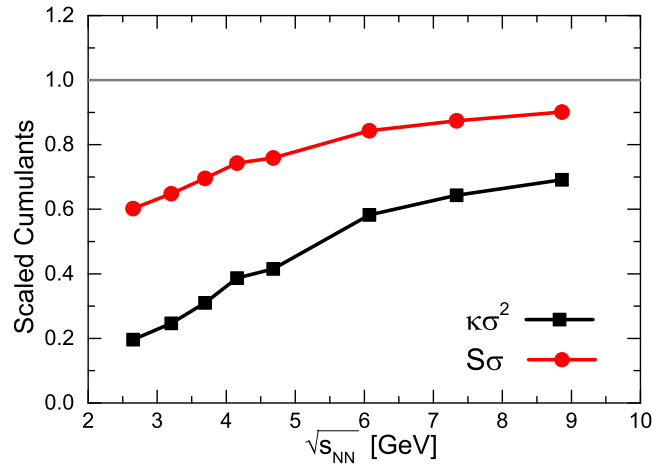}
\caption{\label{f:skkurt} 
Analytic results for the skewness and kurtosis for Au+Au collisions in the energy range 
where antiproton production is not important.}
\end{figure}

Figure \ref{f:skkurt} shows the results for the beam energy dependence of the scaled skewness and kurtosis,
using as input measured mid-rapidity yields of protons and deuterons. We observe a clear deviation from unity for 
all beam energies under consideration. 

\section{Results including anti-protons}
\begin{figure}[t]
\includegraphics[width=0.5\textwidth]{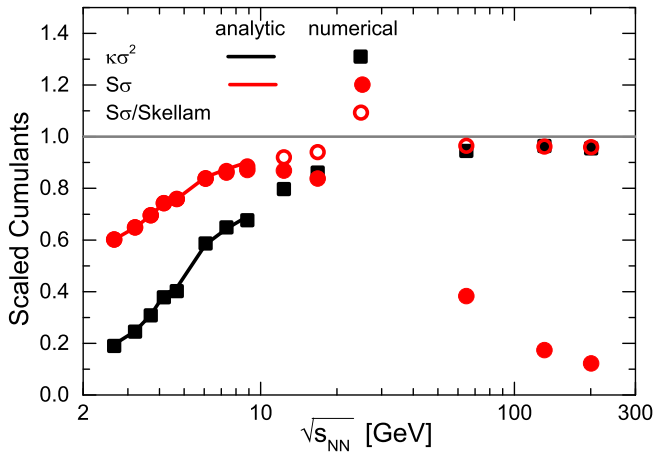}
\caption{\label{fig:fluc} The quantities $S \sigma$ and $\kappa \sigma^2$ as  functions of energy for central Au+Au reactions over a broad range of energies. Compared are the numerical and analytic results. We clearly observe a drastic deviation from the expectation of a Poisson or Skellam distribution.}
\end{figure}

In order to extend our investigations to beam energies higher than $\sqrt{s_{NN}}\approx 10$ GeV, we need to include the effects of anti-protons and anti-deuterons into our study. This extension of our approach is straightforward.
As the measured mean proton and anti-proton numbers are known we can assume that both the proton and anti-proton number
separately follow a Poisson distribution. From this assumption, the initial proton and anti-proton number can be sampled independently and according to their measured multiplicities, 
as described in the previous section. Since the fluctuations of
both are described by uncorrelated Poisson distributions the fluctuations of the initial net-proton number should be described by a Skellam distribution. Assuming that the coalescence parameter for anti-deuterons is identical to that of deuteron formation we can numerically determine the final net-proton number in a given event as
\begin{equation}
n_{p-\overline{p}}= (n_i-n_d)-(n_{\overline{i}} - n_{\overline{d}}) \mathrm{,}
\end{equation}   
as well as the corresponding net-proton number distributions.
To obtain the mean final net-proton number and its cumulants we perform a numerical sampling of the initial proton and anti-proton number as well as the coresponding (anti-)deuteron numbers in each event. To achieve satisfactory statistics we sample $10^9$ events per beam energy.
The resulting scaled cumulants of the proton number distributions, $S \sigma$ and $\kappa \sigma^2$, are shown in Fig.~\ref{fig:fluc} as functions of energy for central Au+Au reactions over a broad range of energies. We compare our numerical results including anti-protons and anti-deuterons 
with the analytical results obtained in the previous section. For the lower SPS energies only small variations are observed.
We clearly confirm the drastic deviation from the expectation of a Poisson or Skellam distribution. The $\kappa \sigma^2$ and $S \sigma$ show strong deviations from unity as the energy decreases and the deuteron fraction increases. This indicates that the formation of deuterons after the kinetic freeze-out may lead to substantial modifications of the fluctuation observable considered here.

\section{Summary}
In the light of the ongoing debate on the interpretation of the recently and currently measured data on fluctuations, we have explored the influence of deuteron formation on fluctuation observables. We employed a standard approach and used that the deuteron formation probability is proportional to the square of the proton yield. This non-linear coupling of the proton and deuteron yields results in a substantial modification of the fluctuation patterns towards low energies: a strong reduction of the $\kappa \sigma^2$ and $S \sigma /$Skellam observable
has been found.

\begin{acknowledgments}
The authors would like to thank R. Holzmann for stimulating this analysis of deuteron formation.
The computational resources were provided by the LOEWE-CSC. This work was supported by HIC for FAIR within the Hessian LOEWE initiative.
ZF and BT acknowledge partial support by grant No's 
APVV-0050-11, VEGA 1/0469/15 (Slovakia) and BT acknowledges 
M\v{S}MT grants LG13031 and LG15001 (Czech Republic).
ZF thanks the DAAD for the support during the stay at the 
Frankfurt Institute for Advanced Studies. 
\end{acknowledgments}


\end{document}